\documentclass[conference]{IEEEtran}

\usepackage{algpseudocode}
\usepackage{algorithm}
\algrenewcommand\alglinenumber[1]{\tiny #1:}

\usepackage{flushend}

\usepackage{graphicx}

\ifCLASSOPTIONcompsoc 
\usepackage[caption=false,font=normalsize,labelfont=sf,textfont=sf]{subfig}
\else
\usepackage[caption=false,font=footnotesize]{subfig}
\fi

\usepackage{wrapfig}

\usepackage{booktabs}

\usepackage{alltt}
\usepackage[flushleft]{threeparttable}

\algnewcommand\algorithmicswitch{\textbf{switch}}
\algnewcommand\algorithmiccase{\textbf{case}}
\algnewcommand\algorithmicassert{\texttt{assert}}

\algdef{SE}[SWITCH]{Switch}{EndSwitch}[1]{\algorithmicswitch\ #1\ \algorithmicdo}{\algorithmicend\ \algorithmicswitch}%
\algdef{SE}[CASE]{Case}{EndCase}[1]{\algorithmiccase\ #1}{\algorithmicend\ \algorithmiccase}%
\algtext*{EndSwitch}%
\algtext*{EndCase}%

\usepackage{cite}

\usepackage{textcomp}

\usepackage{multirow}
\usepackage{hhline}

\usepackage{algorithm}
\usepackage{algpseudocode}

\usepackage{url}
\urldef{\mailsa}\path|{suejb.memeti, sabri.pllana}@lnu.se  |

\begin{document}
%

\title{Analyzing large-scale DNA Sequences on Multi-core Architectures}

\author{\IEEEauthorblockN{Suejb Memeti and
		Sabri Pllana\\}
	\IEEEauthorblockA{Department of Computer Science,
		Linnaeus University\\
		351 95 V\"{a}xj\"{o}, Sweden\\
		\{suejb.memeti, sabri.pllana\}@lnu.se}}

\IEEEspecialpapernotice{\footnotesize(CSE-2015, \copyright IEEE)}

\maketitle


\begin{abstract}
Rapid analysis of DNA sequences is important in preventing the evolution of different viruses and bacteria during an early phase, early diagnosis of genetic predispositions to certain diseases (cancer, cardiovascular diseases), and in DNA forensics. However, real-world DNA sequences may comprise several Gigabytes and the process of DNA analysis demands adequate computational resources to be completed within a reasonable time.  
In this paper we present a scalable approach for parallel DNA analysis that is based on Finite Automata, and which is suitable for analyzing very large DNA segments. We evaluate our approach for real-world DNA segments of mouse (2.7GB), cat (2.4GB), dog (2.4GB), chicken (1GB), human (3.2GB) and turkey (0.2GB). Experimental results on a dual-socket shared-memory system with 24 physical cores show speedups of up to 17.6$\times$. Our approach is up to $3\times$ faster than a pattern-based parallel approach that uses the RE2 library. 

\end{abstract}

\section{Introduction} \label{introduction}
The need for high performance computational biology has emerged as a result of fast growth in biological information, the complexity of interactions that underlie many processes in biology, as well as the diversity and the interconnectedness of organisms at the molecular level \cite{Editorial}. These biological information are accumulated via different techniques, however they require adequate analysis and processing to extract useful information that make the results evident. 

According to Benson et al. \cite{benson2013genbank} the number of Deoxyribonucleic Acid (DNA) sequences and nucleotide bases in these sequences is growing exponentially, doubling every 18 months. As these data are collected, motif search and DNA sequencing are just some examples among many for analytics of Next Gen Sequencing Analysis.

A DNA sequence contains specific genetic instructions, which make the living organisms function properly. In a DNA strand there are four bases of nucleotides: A-adenine, C-cytosine, G-guanine and T-thymine. DNA analysis is important for discovery of differences and similarities of organisms and exploration of the evolutionary relationship between them. This process often requires comparisons of the corresponding DNA sequences, for example, checking whether one sequence is a subsequence of another, or comparing the occurrences of specific $k$-mers in the corresponding DNA sequences. In computational biology $k$-mers refer to all the possible sub-strings (sub-sequences) of length $k$ of a DNA sequence. They have an important role during sequence assembly and can be used in sequence alignment as well. 

Analyzing DNA sequences within a reasonable time is important for domain scientists to study various phenomenons, such as the evolution of viruses and bacteria during an early phase \cite{Mellaman_NGS}, or diagnosis of genetic predispositions to certain diseases.

Modern parallel computing systems promise to provide the capabilities to cope with the DNA analysis processing requirements. Existing approaches use both hardware and software to accelerate regular expression matching. 
The hardware based approaches (such as \cite{Beneson_FSA, Reif_FSA}) are faster, but less flexible and more expensive, whereas software based acceleration techniques are flexible in terms of updating or adding new patterns \cite{SoewitoW07}. Recently different software based DNA analysis techniques designed for multi-core systems have been proposed \cite{Herath,arudchutha2014string,MarcaisK11-jellyfish,nordstrom2013mutation,DrewsLW10,ChaconMEH13}. 

In this paper, we will first explore and discuss the parallelization opportunities of DNA analysis, and thereafter we introduce a parallel algorithm for DNA analysis that is based on Finite Automata. We use a domain decomposition approach for parallelization; in our approach the DNA sequence is split into several chunks, and each chunk is assigned to a thread to perform pattern matching. Our algorithm is optimized to do efficient speculations of the possible initial states for each chunk. Only one regular expression matching (REM) for a chunk is required to be completely performed; the remaining REMs stop when the converging point is reached. A converging point is a state where two or more REM starting from different states meet after the same number of symbols is read. Furthermore, we use a memory efficient data structure that saves the necessary information to count and highlights the $k$-mers. Experiments with real-world DNA segments (for human and various animals) on a dual socket shared-memory system with 48 threads show significant speedups compared to the sequential version (up to 17.6x). The implementation of our algorithm is up to 3x faster than a pattern-based algorithm implemented using the RE2 library \cite{RE2_Library}. Major contributions of this paper include:
\begin{itemize}
	\item a parallel algorithm for DNA analysis that is based on Finite Automata;
	\item empirical evaluation of our algorithm with real-world DNA segments of mouse (2.7GB), cat (2.4GB), dog (2.4GB), chicken (1GB), human (3.2GB) and turkey (0.2GB);
	\item a comparison of our algorithm with a pattern-based algorithm implementation that uses RE2 library.
\end{itemize}

The rest of the paper is organized as follows. Section \ref{methodology} provides background information on pattern matching, whereas Section \ref{our-approach} presents our algorithm for counting and extracting $k$-mers in a DNA sequence. Section \ref{exp_evaluation} presents the experimental setup and discusses the experimental results. The work described in this paper is compared and contrasted to the related work in Section \ref{related_work}. Section \ref{summary_future_work} provides a summary of our work.



\section{Regular Expression Matching (REM) with Finite Automata (FA)}
\label{methodology}



Regular expression matching verifies whether a pattern is present in a string. REM is commonly used for determining the locations of a pattern within a sequence of tokens, in search and replace functions, or to highlight important information out of a huge data set. In the context of computational biology, pattern matching is used for analyzing and processing biological information in order to extract the useful parts of the data and make them evident. 
The formal definition of the REM is as follows: the input text is an array $T[1..n]$ where $n$ is the length of the input, and pattern $P[1..m]$ where the length of the pattern $m \leq n$. The alphabet $\sum$ defines the possible characters of the input string. 

A Finite Automaton (FA) is a machine for processing information by scanning the input text $T$ in order to find the occurrences of the pattern $P$. A formal definition of the FA is as follows: FA is a quintuple of ($Q, \sum, \delta, q_0, F$), where $Q$ is the finite set of states, $\sum$ is the finite alphabet, $\delta$ is the transition function $Q \times \sum \longrightarrow Q$, $q_0$ is the start state and $F$ is the distinguished set of final states.

A well known algorithm for multiple pattern matching is the Aho-Corasick algorithm. It is able to match any occurrences (including the overlapped ones) of multiple patterns linearly to the size of the input string. It examines each character of the input string only once. It builds an automaton by creating states and transitions corresponding to these states. It adds failure transitions when there is no regular transition leaving from the current state on a particular character, which makes it possible to match multiple and overlapping occurrences of the patterns. Furthermore, this algorithm is capable of delivering input-independent performance if implemented efficiently in parallel systems, which is a reason why we use this algorithm as basis of our work.


\section {Design and Implementation of a DNA Analysis algorithm}
\label{our-approach}
In this section we first provide the details about the outline of our algorithm. Thereafter we discuss the most important implementation aspects to achieve a scalable algorithm for counting and extracting specific $k$-mers from a large DNA sequence.

\subsection{Our algorithm for counting and extracting the location of $k$-mers ($k$-mers CoEx)}
\label{coex}
\begin{figure}[ht]

	\centering
	\subfloat[Input-based approach]{
		\includegraphics[width=2.5in]{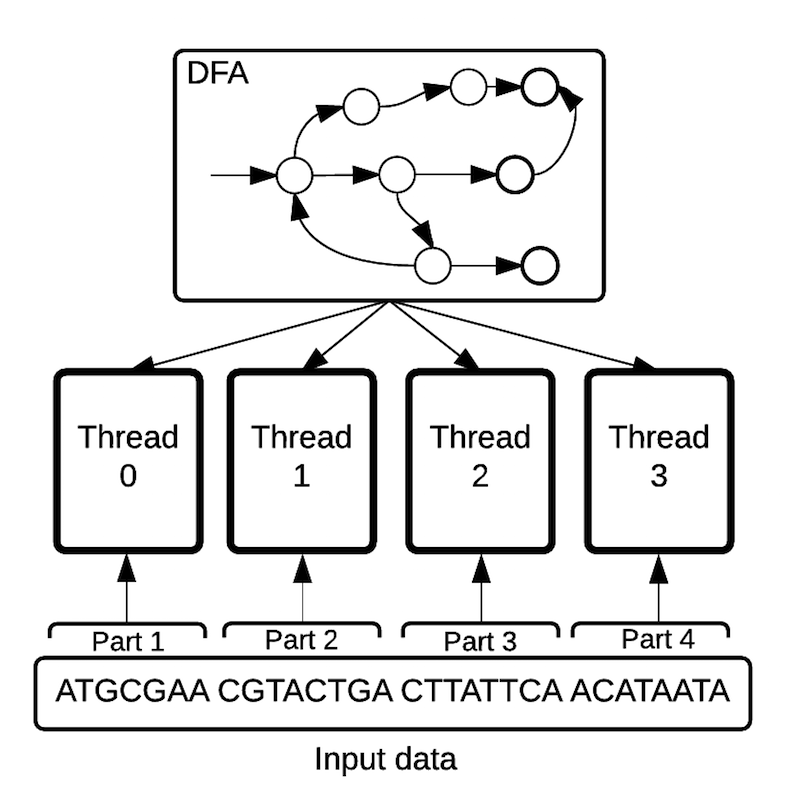}
		\label{fig:split}
	}
	\hfil
	\subfloat[Pattern-based approach]{
		\includegraphics[width=2.5in]{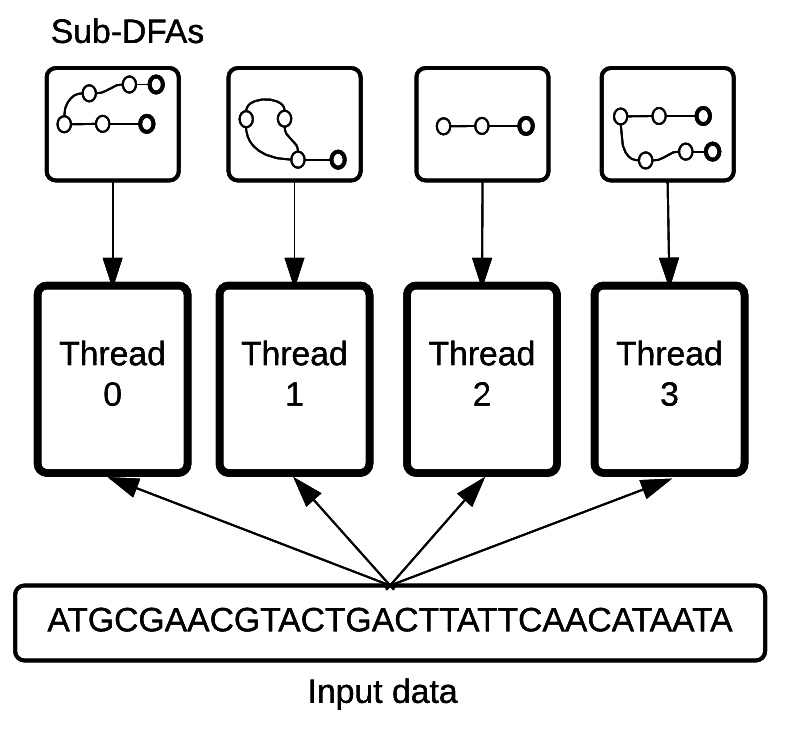}
		\label{fig:simd}
	}
	\caption{Load balancing using Input and Pattern partitioning approach.}
	\label{fig:work-partitioning}
\end{figure}

Figure \ref{fig:work-partitioning} depicts two possible ways for parallel execution of regular expression matching for bio-computing applications: (a) \emph{input-based approach} that splits the input string into smaller chunks and processes them in separate threads and (b) \emph{pattern-based approach} that splits the patterns in sub-patterns, creating separate state machines for each of them and processing the same input string with each different machine \cite{Herath}. 

Our algorithm uses the input-based approach. The challenge of this approach is determining the initial state for each chunk. Finding the correct starting state for each chunk is important for finding the occurrences of the patterns that appear in the crossing border. Other researchers use different ways of finding the initial states, for instance Luchaup et al. \cite{LuchaupSEJ11} use speculation to find the initial state based on the most visited states, Devi and Rajagopalan \cite{Devi} use an index based technique, Chacon et al. \cite{ChaconMEH13} use Suffix-Arrays, Villa et al. \cite{VillaCM09}, uses the pattern length overlapping approach. 

Our way of determining the possible initial states is as follows: (1) find the set of source states ($L$) for the first element of the sub-input mapped to the running thread ($T_{n}$); (2) find the set of destination states ($S$) for the last character of the sub-input mapped to the previous thread ($T_{n-1}$); (3) find the intersection of $S$ and $L$ ($S \cap L$), which is the set of possible initial states \cite{parem}. The first thread ($T_{0}$) always starts from the initial state $q_{0}$. Each thread is responsible for finding the set of possible initial states, and for each state of this set a regular expression matching is performed. When all threads have finished their job, the results are joined by a binary reduction, which connects the last visited state of $T_n$ to the first visited state of $T_{n+1}$.

This method provides very good results for sparse transition tables and good performance for dense matrices for DFAs with relatively small number of states. However, in a DFA with large number of states this method seems to be less efficient. This happens because one thread may be responsible to perform multiple REM for the same input, due to multiple possible initial states.
To reduce the operations required for each thread to perform the REM starting from different states, further optimizations are needed.

While investigating the REM using the modified Aho-Corasick DFA representation, we noticed that the result converges after several symbols are read (in our experiments, 10 is the max number of steps required to find the converging point). A converging point is a state where two or more REM starting from different states meet after the same number of symbols are examined. This insight allows us to significantly minimize the execution cost required to perform the REM starting from each possible initial state. The details about the process of the convergence are given in the next Section. 

\subsection{Implementation Aspects} 

\begin{figure}
	\begin{center}
		\includegraphics[width=3in]{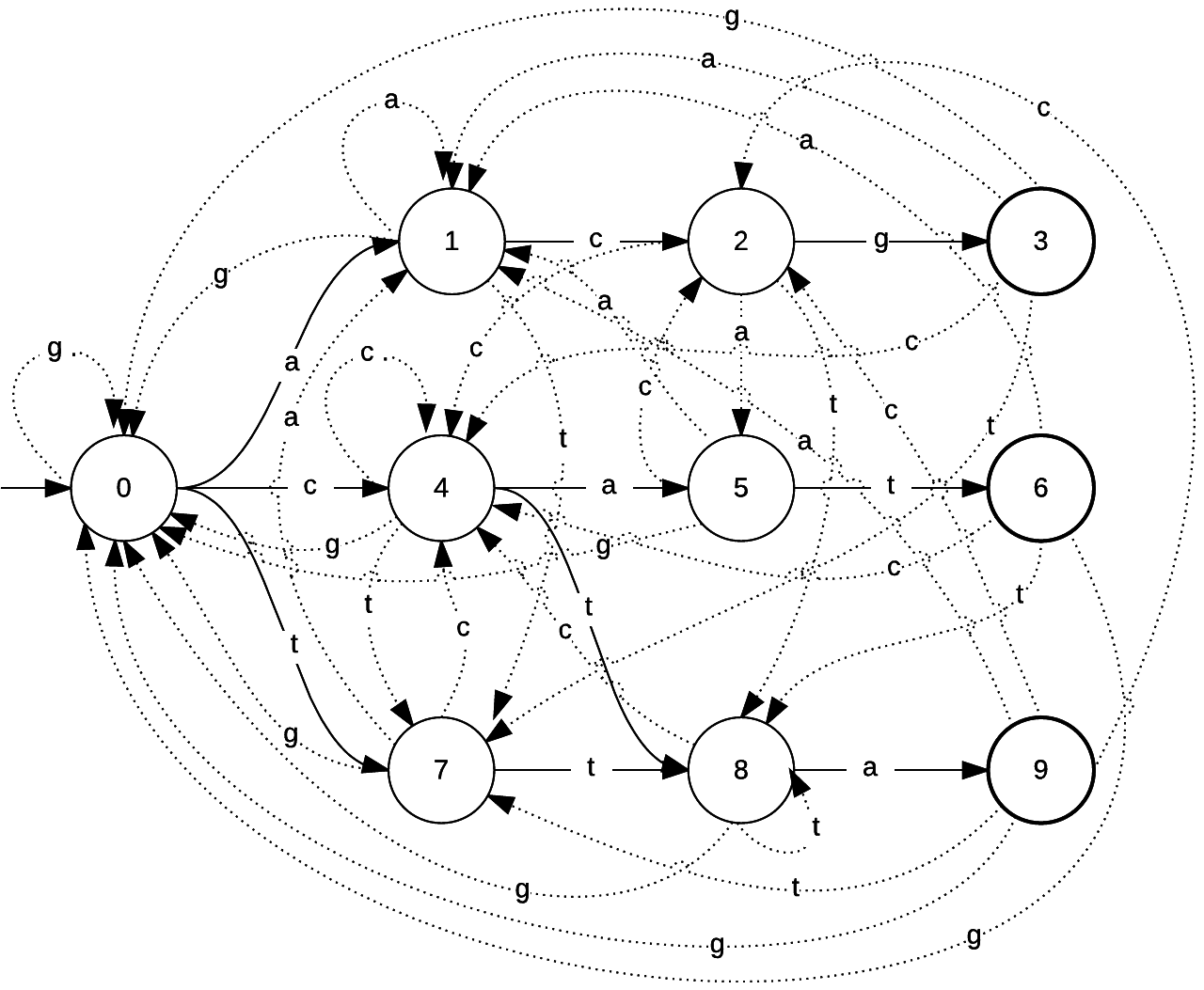}
	\end{center}

	\caption{K-mers CoEx finite state automaton for matching the patterns: $"acg"$, $"cat"$, $"cta"$ and $"tta"$}
	\label{fig:automaton}
\end{figure}

In this section we will explain the implementation details of our algorithm, including the process of building the DFA, splitting the input among the available threads, finding the set of possible initial states for each chunk assigned to a thread, running the REM for the first state in this set and the process of finding the \emph{converging point}.
A reference of each process to the corresponding lines of codes in Algorithm \ref{alg:k-mers-coex} will be provided. 

Furthermore, we define $q_i$ as the state of the automaton where $i$ is the state id, $E_i$ as a sub-pattern of the selected patterns, $R_i$ as the process of performing an REM starting from the item at index $i$ of the set of possible initial states. The values used in the \emph{switch-case} (Line \ref{alg:k-mers-coes:switch-start} - \ref{alg:k-mers-coes:switch-end}) (ex. 122, 127, 128...) determine that a specific sub-pattern ($E_i$) has been matched.

\subsubsection{Building the Deterministic Finite Automaton}

The AC algorithm with failure transitions has a drawback due to the non-deterministic transitions for a single input character. Figure \ref{fig:automaton} illustrates our solution to eliminate the failure transitions by adding the right transition (indicated by dashed lines) for each state. Having a valid transition for each possible character to another state in the automaton, guarantees that for each symbol the same amount of operations will be performed. The example automaton shown on Figure \ref{fig:automaton} is able to match the following patterns: $"acg"$, $"cat"$, $"cta"$ and $"tta"$. For example, if we read string $"ac"$ we reach state $q_{2}$, and when $"a"$, $"c"$ or $"t"$ is read we know exactly that state  $q_{6}$, $q_{5}$ or $q_{10}$ is next, respectively.

\subsubsection{Splitting the input and finding the possible starting states (PSS)}
The process of splitting the input among the available threads is depicted in Table \ref{table:input_split}.a. This is a straight forward step, where the input length is divided by the number of available threads (Line \ref{alg:k-mers-coex:start-split}-\ref{alg:k-mers-coex:end-split}). The chunks are assigned to the threads consecutively based on the thread IDs. 

The pseudo-code of the process of finding the PSS (See Section \ref{coex}) is shown on Algorithm \ref{alg:k-mers-coex} Line \ref{alg:k-mers-coex:get_PSS_0}. When the PSS are determined the thread performs an REM starting from each item of PSS (Line \ref{alg:k-mers-coex:rem-start}). Table \ref{table:input_split}.b,c depicts the REM process starting from each item of PSS; each table corresponds to a thread. For example, the first thread (see Table \ref{table:input_split}.b) initiates the REM starting from the following PSS: $q_3$, $q_{10}$, $q_{13}$, and $q_{139}$.

For every $R_i$ (REM starting from the $PSS$ at index $i$) a $CR$ structure is created and stored in the $results$. The $CR$ structure stores the \emph{initial state}, \emph{last state} and the total number of occurrences for each of the sub-patterns (Line \ref{alg:k-mers-coex:cr_start} - \ref{alg:k-mers-coex:cr_end}).

\begin{algorithm}
	\caption{$k$-mers CoEx}\label{alg:k-mers-coex}
	\scriptsize
	\begin{algorithmic}[1]
		\Statex{\textbf{Input}: transition table $dfa$; set of final states $F$; input string $I$}
		\Statex{\textbf{Output}: list of $CR$ $results$}
		\Procedure{kcoex}{$dfa, F, I$}
		\State $steps = 10$\Comment{The max number of steps required to converge}
		\State $result = list<list<CR>>$ \Comment{Stores the final states for each thread} \label{alg:k-mers-coex:global-data-structure}
		\For{$T_0 ... T_n$} in parallel \Comment{$T$ - Thread, $n$ - total number of threads}
		\State{$start\_position = t\_i * (I.length / n)$} \Comment{$t\_i = thread\_id$} \label{alg:k-mers-coex:start-split}
		\State{$SI = substring(start\_position, I.length / n)$} \Comment{SI - sub input} \label{alg:k-mers-coex:end-split}
		\If{$t_i ~!= 0$}
		\State{$PSS = get\_possible\_starting\_states(I[start\_position],$ $I[start\_position-1])$} \label{alg:k-mers-coex:get_PSS}
		\Else
		\State{$PSS = [0]$} \label{alg:k-mers-coex:get_PSS_0}
		\EndIf
		\State{$fr\_list = list<FR>$} \label{alg:k-mers-coex:fr} 
		\State{$psi\_i = 0$}
		\For{$int ~cs ~in~ PSS$} \label{alg:k-mers-coex:rem-start}
		\State{$CR ~ cr$} \Comment{stores the init state, last state, and total number of occurrences for each subexpression} \label{alg:k-mers-coex:cr}
		\State{$char\_i = 0$}
		\For{$char ~c ~ in ~ SI$}
		\If{ $char\_i == 0$ }
		\State {$cr.init\_state = cs$} 
		\ElsIf{$char\_i == SI.length - 1$}
		\State {$cr.last\_state = dfa[cs][c]$} 
		\EndIf 
		\State{$cs = dfa[cs][c]$}
		\If{$cs ~ in ~ F$}
		\Switch{$cs$} \label{alg:k-mers-coes:switch-start}
		\Case{$117$}
		\State{$cr.final\_states[0] ++$}	\Comment{$agggtaaa \mid tttaccct$ is found}	
		\EndCase
		\Case{$122$ or $128$}
		\State{$cr.final\_states[1] ++$} \Comment{$(c|g|t)gggtaaa \mid tttaccc(a|c|g)$ is found}
		\EndCase
		\State{...}
		\EndSwitch \label{alg:k-mers-coes:switch-end}

		\label{alg:k-mers-coex:save-to-global}
		\EndIf
		\If{$psi\_i == 0 ~and~ char\_i \leq steps$} \label{alg:k-mers-coex:fullrun-start}
		\State {$FR ~fr$}
		\State{$fr.current\_state = cs$} 
		\State{$fr.final\_states[0] = cr.final\_states[0]$} 
		\State{...}
		\State{$fr.final\_states[8] = cr.final\_states[8]$} 
		\State{$fr\_list.add(fr)$}
		\label{fullrun-end}
		\ElsIf {$psi\_i > 0 ~and~ fr\_list[char\_i].current\_state == cs$} \label{alg:k-mers-coex:converge-point} \Comment{check for convergence}
		\State{$cr.final\_states[0] ~+=~results[t\_i][0].final\_states[0] - fr\_list[char\_i].final\_states[0]$}
		\State{...}
		\label{alg:k-mers-coex:converge-point-calc}
		\State {\textbf{break} }
		\EndIf 
		\EndFor
		\State{$results[t\_i].add(cr)$}
		\EndFor
		
		\EndFor
		\EndProcedure
		\Statex{}
		\Statex{\textbf{Input}: transition table $dfa$; the first character of the input mapped to $T_n$ (current thread) $first\_char$; the last character of the input mapped to $T_{n-1}$ (previous thread) $last\_char$}
		\Statex{\textbf{Output}: list of states}
		\Procedure{get\_possible\_starting\_states}{$dfa, first\_char, last\_char$} \label{alg:kmers-coes:psi_start}
		\State{$S = L = list<q>$} \Comment {$q$ - state of the DFA}
		\For{$q_0 ... q_n$}
		\If{ $dfa[q_i][first\_char] \in Q $} \Comment{$Q$-list of states}
		\State {$S_i=q_i$}
		\EndIf 
		
		\If{ $dfa[q_i][last\_char] \in Q $}
		\State {$L_i=dfa[q_i][last\_char]$}
		\EndIf
		\EndFor
		\State {\textbf{return} $S \cap L$}
		\EndProcedure \label{alg:kmers-coes:psi_end}
		\Statex{}
		\State{\textbf{struct} $CR$\{} \label{alg:k-mers-coex:cr_start}
		\State{\ \ \ \ \ \textbf{int} $init\_state$}
		\State{\ \ \ \ \ \textbf{int} $last\_state$}
		\State{\ \ \ \ \ \textbf{int} $final\_states[9]$} \Comment{Stores the number of occurrences for each sub-expression (E1...E9)}
		\State{\}} \label{alg:k-mers-coex:cr_end}
		\Statex{}
		\State{\textbf{struct} $FR$\{} \label{alg:k-mers-coex:fr_start}
		\State{\ \ \ \ \ \textbf{int} $current\_state$}
		\State{\ \ \ \ \ \textbf{int} $final\_states[9]$}
		\State{\}} \label{alg:k-mers-coex:fr_end}
	\end{algorithmic}
\end{algorithm}

\begin{table}[!h]	
		\renewcommand{\arraystretch}{1.4}
		\setlength{\tabcolsep}{4pt}
		\caption{The process of splitting the input, performing the REM starting from each possible initial state, and the process of convergence
			}
		\label{table:input_split}
		\begin{tabular}{lllllllllllllll}
			\multicolumn{15}{l}{a) Splitting the input string into chunks} \\ 
			
			& & & $_0$ & $_1$ & $_2$ & $_3$ & $_4$ & $_5$ & $_6$ & $_7$ & $_8$ & $_9$ & &  \\ 
			\cline{2-2} \cline{4-14}

			& \multicolumn{1}{|l|}{$T_1$} & \multicolumn{1}{l|}{$\rightarrow$} & \multicolumn{1}{l|}{G} & \multicolumn{1}{l|}{T} & \multicolumn{1}{l|}{G} & \multicolumn{1}{l|}{A} & \multicolumn{1}{l|}{G} & \multicolumn{1}{l|}{C} & \multicolumn{1}{l|}{C} & \multicolumn{1}{l|}{G} & \multicolumn{1}{l|}{A} & \multicolumn{1}{l|}{G} & \multicolumn{1}{l|}{\ldots} & chunk 1  \\ 
			\cline{2-2} \cline{4-14}

			& \multicolumn{1}{|l|}{$T_2$} & \multicolumn{1}{l|}{$\rightarrow$} & \multicolumn{1}{l|}{C} & \multicolumn{1}{l|}{C} & \multicolumn{1}{l|}{C} & \multicolumn{1}{l|}{T} & \multicolumn{1}{l|}{A} & \multicolumn{1}{l|}{T} & \multicolumn{1}{l|}{A} & \multicolumn{1}{l|}{C} & \multicolumn{1}{l|}{G} & \multicolumn{1}{l|}{A} & \multicolumn{1}{l|}{\ldots} & chunk 2   \\ 
			\cline{2-2} \cline{4-14}
			\multicolumn{15}{l}{} \\
		\end{tabular}

		\begin{tabular}{llllllllllllll}
			\multicolumn{14}{l}{b) REM for chunk 1} \\ 
			\cline{2-2} \cline{4-14} 
			
			\multicolumn{1}{l|}{\multirow{4}{*}{\rotatebox[origin=c]{90}{Initial state}}} & \multicolumn{1}{l|}{1}  & \multicolumn{1}{l|}{$\rightarrow$} & \multicolumn{1}{l|}{7}   & \multicolumn{1}{l|}{21} & \multicolumn{1}{l|}{31}  & \multicolumn{1}{l|}{1} & \multicolumn{1}{l|}{7}   & \multicolumn{1}{l|}{19} & \multicolumn{1}{l|}{2} & \multicolumn{1}{l|}{9} & \multicolumn{1}{l|}{1} & \multicolumn{1}{l|}{7} & \multicolumn{1}{l|}{\dots}\\ 
			\cline{2-2} \cline{4-14} 
			
			\multicolumn{1}{l|}{} & \multicolumn{1}{l|}{6}  & \multicolumn{1}{l|}{$\rightarrow$} & \multicolumn{1}{l|}{16}  & \multicolumn{1}{l|}{21} & \multicolumn{1}{l|}{\dots} & & & & & & & &  \\
			\cline{2-2} \cline{4-8}
			
			\multicolumn{1}{l|}{} & \multicolumn{1}{l|}{18} & \multicolumn{1}{l|}{$\rightarrow$} & \multicolumn{1}{l|}{32}  & \multicolumn{1}{l|}{48} & \multicolumn{1}{l|}{13}  & \multicolumn{1}{l|}{1} & \multicolumn{1}{l|}{\dots} & & & & & & \\ 
			\cline{2-2} \cline{4-8}
			
			\multicolumn{1}{l|}{} & \multicolumn{1}{l|}{43} & \multicolumn{1}{l|}{$\rightarrow$} & \multicolumn{1}{l|}{63}  & \multicolumn{1}{l|}{21} & \multicolumn{1}{l|}{\dots} & & & & & & & & \\ 
			\cline{2-2} \cline{4-6}
			
			\multicolumn{14}{l}{} \\
		\end{tabular}
		
		\begin{tabular}{llllllllllllll}
			\multicolumn{14}{l}{c) REM for chunk 2} \\ 
			\cline{2-2} \cline{4-14} 
			
			\multicolumn{1}{l|}{\multirow{4}{*}{\rotatebox[origin=c]{90}{Initial state}}} & \multicolumn{1}{l|}{3}  & \multicolumn{1}{l|}{$\rightarrow$} & \multicolumn{1}{l|}{2}  & \multicolumn{1}{l|}{2}   & \multicolumn{1}{l|}{2}   & \multicolumn{1}{l|}{10}  & \multicolumn{1}{l|}{11} & \multicolumn{1}{l|}{130} & \multicolumn{1}{l|}{40} & \multicolumn{1}{l|}{60} & \multicolumn{1}{l|}{15} & \multicolumn{1}{l|}{1}   & \multicolumn{1}{l|}{\dots} \\ 
			\cline{2-2} \cline{4-14} 
			
			\multicolumn{1}{l|}{} & \multicolumn{1}{l|}{10} & \multicolumn{1}{l|}{$\rightarrow$} & \multicolumn{1}{l|}{19} & \multicolumn{1}{l|}{2}   & \multicolumn{1}{l|}{\dots} & & & & & & & &\\ 
			\cline{2-2} \cline{4-13}
			
			\multicolumn{1}{l|}{} & \multicolumn{1}{l|}{44} & \multicolumn{1}{l|}{$\rightarrow$} & \multicolumn{1}{l|}{67} & \multicolumn{1}{l|}{90}  & \multicolumn{1}{l|}{112} & \multicolumn{1}{l|}{129} & \multicolumn{1}{l|}{1}  & \multicolumn{1}{l|}{8}   & \multicolumn{1}{l|}{11} & \multicolumn{1}{l|}{5}  & \multicolumn{1}{l|}{15} & \multicolumn{1}{l|}{\dots} & \\
			\cline{2-2} \cline{4-13}
			
			\multicolumn{1}{l|}{} & \multicolumn{1}{l|}{63} & \multicolumn{1}{l|}{$\rightarrow$} & \multicolumn{1}{l|}{84} & \multicolumn{1}{l|}{105} & \multicolumn{1}{l|}{2}   & \multicolumn{1}{l|}{\dots} & & & & & & &\\ 
			\cline{2-2} \cline{4-7}
		\end{tabular}
\end{table}

\begin{table}

		\setlength{\tabcolsep}{5.7pt}
		\caption{A tabular representation of the $fr\_list$ (Line  \ref{alg:k-mers-coex:fr})}
		\label{table:fr_structure}
		
		\begin{tabular}{lllllllllll}
			\toprule
			\emph{Step} & \emph{CS} & $E_1$ & $E_2$ & $E_3$ & $E_4$ & $E_5$ & $E_6$ & $E_7$ & $E_8$ & $E_9$\\ \midrule
			\emph{1} & 67  & 0 & 0 & 0 & 0 & 0 & 0 & 0 & 0 & 0 \\ \hline
			& & & & & \multicolumn{6}{l}{...} \\ \hline
			\emph{4} & 129 & 0 & 0 & 0 & 0 & 0 & 1 & 0 & 0 & 0 \\ \bottomrule
		\end{tabular}
\end{table}

\begin{table}	
		\begin{threeparttable}
			\renewcommand{\arraystretch}{1.1}
			\setlength{\tabcolsep}{5pt}
			\caption{The tabular representation of the $results$ (Line \ref{alg:k-mers-coex:global-data-structure})}
			\label{table:data_structure}
			
			\begin{tabular}{llllllllllll}
				\toprule
				$T$& $Q_0$ & $Q_n$ & $E_1$ & $E_2$ & $E_3$ & $E_4$ & $E_5$ & $E_6$ & $E_7$ & $E_8$ & $E_9$ \\ \midrule
				\multirow{2}{*}{1} & 1     & 130  & 0 & 0 & 0 & 1 & 0 & 0 & 0 & 0 & 0 \\
				& 96    & 130  & 0 & 0 & 0 & 1 & 0 & 0 & 0 & 0 & 0 \\ \midrule

				\multirow{2}{*}{2} & 3     & 130  & 0 & 0 & 1 & 0 & 1 & 0 & 2 & 0 & 1 \\ 
				& 44    & 130  & 0 & 0 & 1 & 0 & 1 & 1 & 2 & 0 & 1 \\ \bottomrule
			\end{tabular}
			\begin{tablenotes}
				\small
				\item $T$ - thread index
				\item $Q_0$ - start state, $Q_n$ - end state
				\item $E_1 - E_9$ - sub-expressions
				\item $CS$ - current state
			\end{tablenotes}
		\end{threeparttable}
\end{table}

\subsubsection{Determining the Converging Point}

We investigated that while performing an REM of the same input string starting from different states, the REM converges after a certain number of steps. 
An example of how the convergence happens is depicted in Table \ref{table:input_split}.b,c. For example in Table \ref{table:input_split}.c, the matching of \emph{CCCTATACGA...} (see Table \ref{table:input_split}.a) starting from $q_3$ leads to the following transitions: $q_2 \rightarrow q_2 \rightarrow q_2 \rightarrow q_{10} \rightarrow q_{11} \rightarrow q_{130} \rightarrow q_{40} \rightarrow q_{60} \rightarrow q_{15} \rightarrow q_1 ...$. Matching the same input starting from $q_{10}$ leads to the following transitions: $q_{19} \rightarrow q_2$; no further transitions are needed, because the state at position two for both REMs ($R_0$ and $R_1$) is $q_2$, which indicates the point of convergence.

In order to properly count the number of occurrences of $k$-mers when a converging point is met, we need to store the number of occurrences of $k$-mers for the first $n$ steps while performing $R_0$. For each of the first $n$ examined characters a $FR$ structure is created and stored in the $fr\_list$ (Line \ref{alg:k-mers-coex:fullrun-start}-\ref{fullrun-end}). The $FR$ structure is shown on Algorithm \ref{alg:k-mers-coex} Line \ref{alg:k-mers-coex:fr_start} - \ref{alg:k-mers-coex:fr_end}, which stores the \emph{current state} of the automaton, and the current number of occurrences for each of the sub-patterns. An example of this process is depicted in Table \ref{table:fr_structure}, where each row represents a $FR$ structure. In this example, we assume that $q_{44}$ is the first state in the PSS, which means that the $R_0$ starts from $q_{44}$. After four characters ($CCCT$) are examined, a final state ($q_{129}$ - representing $E_6$, see Table \ref{table:input_split}.e row 3) is reached, therefore the number of current occurrences of $k$-mers for $E6$ becomes 1 (see Table \ref{table:fr_structure}). 

The REM starting from the remaining states of $PSS$ will be performed until the converging point is reached. For instance (see Table \ref{table:input_split}.b), $R_1, R_2~ and ~R_3$ need only 2, 4, 2 characters to be examined, respectively. When the converging point is reached, the total number of final states ($CR.final\_states$) is calculated by adding the total number of states found for $R_{0}$ ($results[t\_i][0].final\_states$) to the $R_{i}$ and subtracting the final states ($fr\_list[char\_i].final\_states$) found of $R_{0}$ at the converging point (Line \ref{alg:k-mers-coex:converge-point-calc}).


\section{Experimental Evaluation} \label{exp_evaluation}

In this section we describe the experimentation environment used for the evaluation of our proposed algorithm and we discuss the obtained performance results.

\subsection{Experimentation Environment}

We have performed experiments on a shared-memory system with two 12-core Intel Xeon processors of the type E5-2695 v2 and 16GB Memory. In total the system has 24 physical cores and each physical core supports two threads (also known as logical cores). We have implemented our algorithm using C++11 programming language and OpenMP. For compilation we used the Intel Compiler icc 15.0.0. In order to address the variability in performance measurements we have repeated each experiment 20 times. 

For our experimental evaluation we have selected data-sets of genomes from the GenBank National Center for Biotechnology Information sequence database \cite{GenBank}: mouse (2.7GB), cat (2.4GB), dog (2.4GB), chicken (1GB), human (3.2GB) and turkey (0.2GB). The information about the data-sets is provided in Table \ref{table:dna-sequences}.

\begin{table}[h]
	\renewcommand{\arraystretch}{1.3}
	\caption{DNA data-sets}
	\label{table:dna-sequences}
	\centering
	\begin{tabular}{lll}
		\toprule
		& \emph{Genome Reference}   	 & \emph{Size (MB)} \\ \midrule
		\emph{Mouse}   & GRCm38.p2        		 & 2830      \\ 
		\emph{Cat}     & Felis\_catus-6.2  	 & 2490      \\ 
		\emph{Dog}     & CanFam3.1          	 & 2440      \\ 
		\emph{Chicken} & Gallus\_gullus-4.0	 & 1060      \\ 
		\emph{Human}   & GRCh38            	 & 3250      \\ 
		\emph{Turkey}  & Meleagris\_gallopavo   & 193       \\ \bottomrule
	\end{tabular}
\end{table}

In our experiments we used the same patterns as in the \emph{regex-dna} benchmark, listed in Table \ref{table:regex} \cite{REGEX_DNA}, which are used to extract and match DNA 8-mers and substitute nucleotides according to standards of International Union of Biochemistry (IUB). In Section \ref{sec:results} we compare the performance of \emph{regex-dna} benchmark with our $k$-mers CoEx algorithm.

\begin{figure*}[!t]
	\centering
	\includegraphics[width=\linewidth]{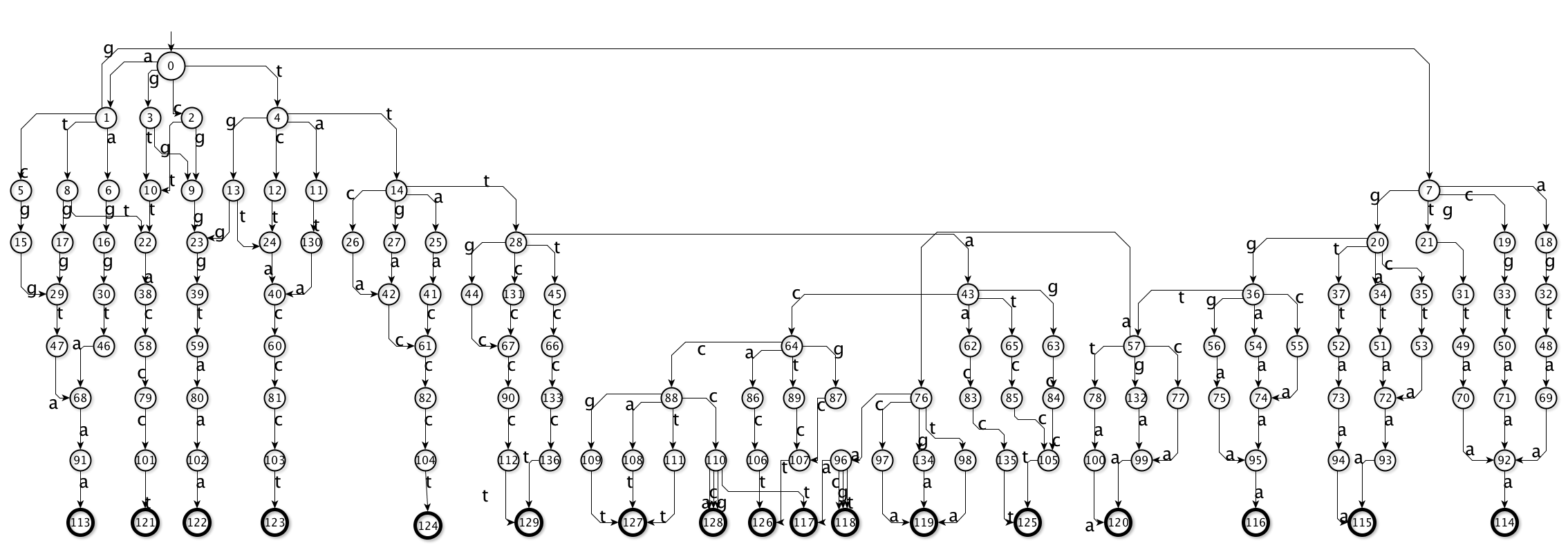}
	\caption{The DFA automaton that matches (counts and extracts the location of $k$-mers) 8-mers from a given DNA sequence. The corresponding regular expression is shown on Table \ref{table:regex}. For simplicity the failure links are omitted from the figure.}
	\label{fig:dna_seq_automaton}
\end{figure*}

The DFA for the given regular expression (Table \ref{table:regex}) was generated using our PaREM tool \cite{parem}. Figure \ref{fig:dna_seq_automaton} depicts the DFA of 137 states, which is able to find the occurrences (including the overlapping ones) of the selected patterns. For simplicity the failure links are omitted from the DFA graph.

\begin{table}[ht]
	\renewcommand{\arraystretch}{1.3}
	\caption{Patterns of the \emph{regex-dna} benchmark. The symbol "$|$" determines the "OR" regex operator}
	\label{table:regex}
	\centering
	\begin{tabular}{ p{0.5cm} p{2.2cm} p{0.2cm} p{2.5cm} }
		\hline 
		$E1$ & $agggtaaa$ & $|$ & $tttaccct$\\
		$E2$ & $(c|g|t)gggtaaa$ & $|$ & $tttaccc(a|c|g)$\\
		$E3$ & $a(a|c|t)ggtaaa$ & $|$ & $tttacc(a|g|t)t$\\
		$E4$ & $ag(a|c|t)gtaaa$ & $|$ & $tttac(a|g|t)ct$\\
		$E5$ & $agg(a|c|t)taaa$ & $|$ & $ttta(a|g|t)cct$\\
		$E6$ & $aggg(a|c|g)aaa$ & $|$ & $ttt(c|g|t)ccct$\\
		$E7$ & $agggt(c|g|t)aa$ & $|$ & $tt(a|c|g)accct$\\
		$E8$ & $agggta(c|g|t)a$ & $|$ & $t(a|c|g)taccct$\\
		$E9$ & $agggtaa(c|g|t)$ & $|$ & $(a|c|g)ttaccct$\\
		\hline
	\end{tabular}
\end{table}

\subsection{Results}
\label{sec:results}
\begin{figure*}[!t]
	\centering
	\includegraphics[width=\linewidth]{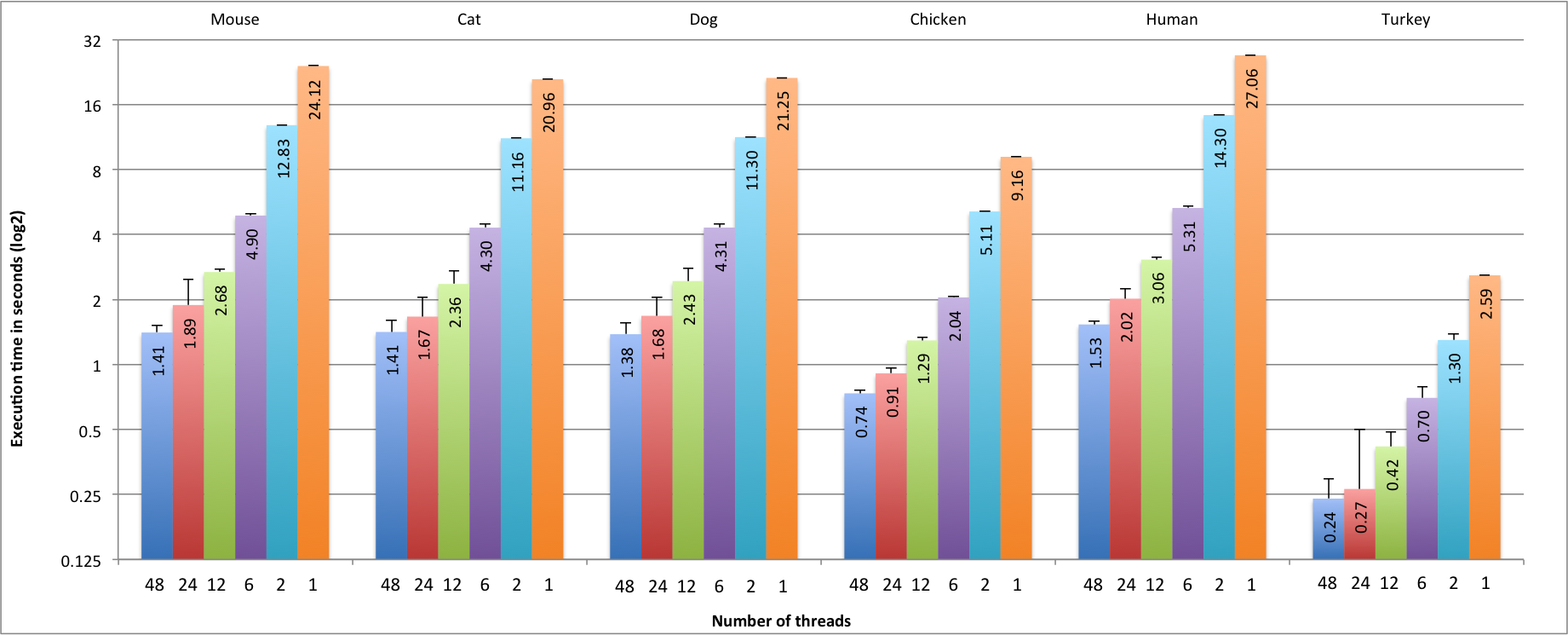}
	\caption{Performance results of our $k$-mers CoEx algorithm for various numbers of threads and data-sets. As input are used six DNA sequences of various lengths: mouse (2.7GB), cat (2.4GB), dog (2.4GB), chicken (1GB), human (3.2GB), and turkey (0.2GB). The experiments are performed by varying the number of threads (2, 6, 12, 24, and 48). The performance measurements for each experiment have been repeated 20 times.}
	\label{fig:executiontime}
\end{figure*}

We first present the performance results of our $k$-mers CoEx algorithm for various problem and machine sizes, and thereafter we compare our algorithm with a pattern-based algorithm implementation that uses RE2 library (known as \emph{regex-dna} benchmark). Figure \ref{fig:executiontime} depicts the execution time in logarithmic scale for each of our selected data-sets and for various numbers of threads \{1,2,6,12,24,48\}. We observe a good scalability of our algorithm as we increase the number of threads or the input size. For example, the analysis of the human's DNA sequence using one thread takes 27 seconds, and by increasing the number of threads to 2, 6, 12, 24 and 48 the execution time reduces to 14.3s, 5.3s, 3s, 2s and 1.5s respectively.

\begin{figure}
		\centering
		\includegraphics[width=\linewidth]{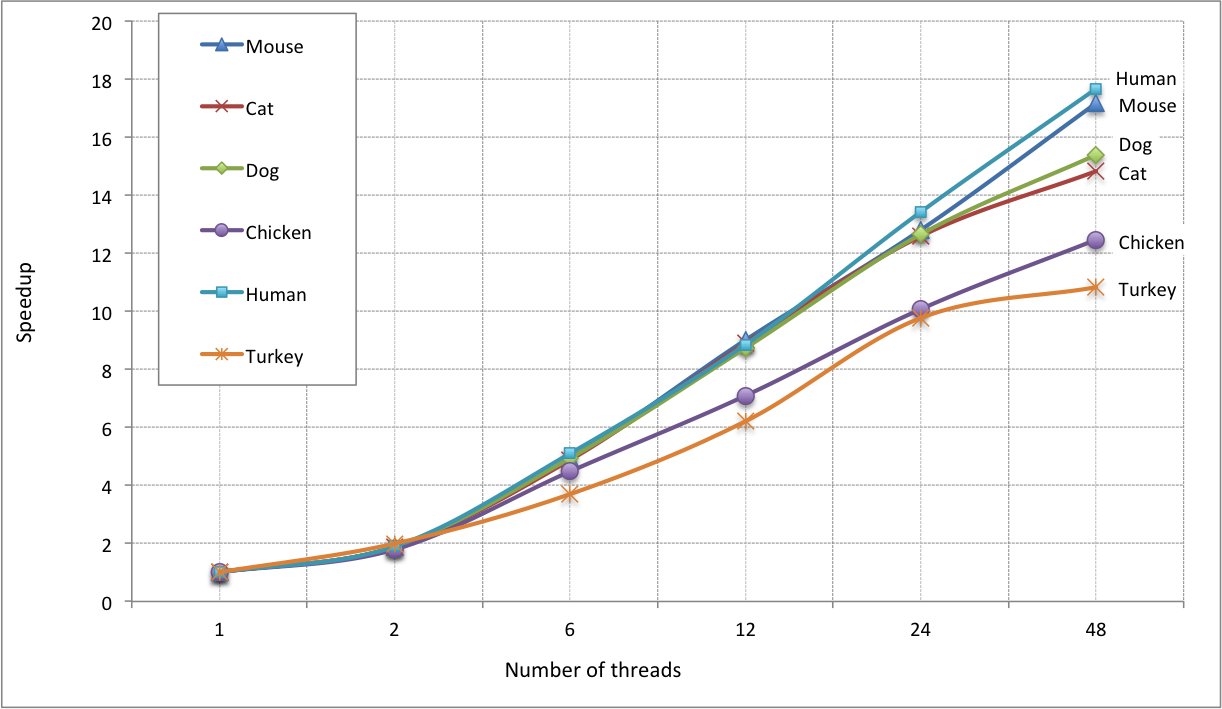}
		\caption{Speedup of our $k$-mers CoEx algorithm implementation.}
		\label{fig:speedup}
\end{figure}
\begin{figure}
		\centering
		\includegraphics[width=\linewidth]{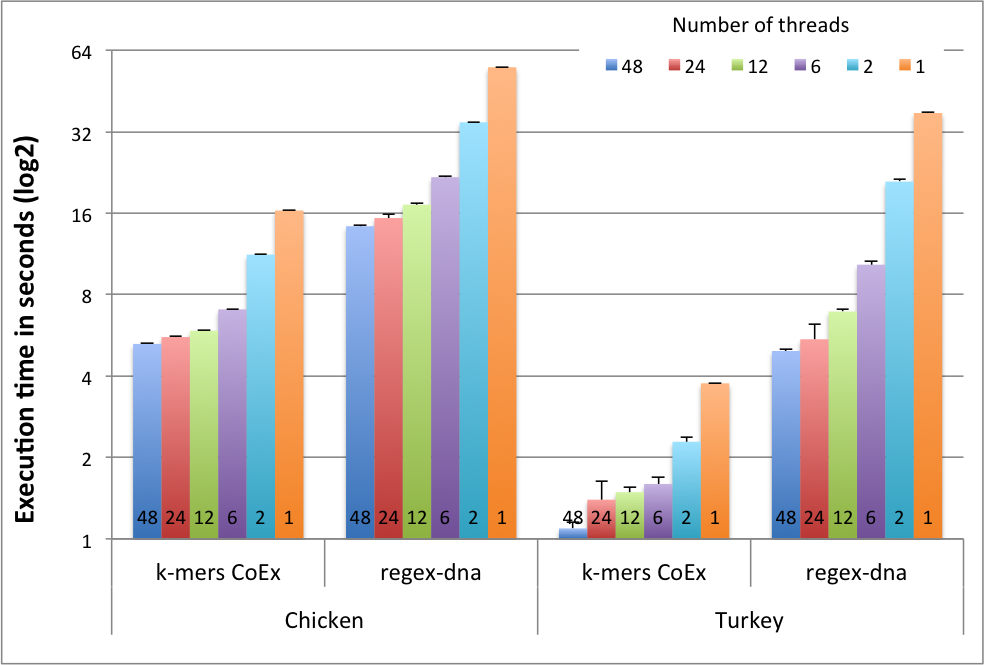}
		\caption{Performance comparison between $k$-mers CoEx and the \emph{regex-dna benchmark} (RE2)}
		\label{fig:comparison}
\end{figure}

Figure \ref{fig:speedup} depicts the obtained speedup of our algorithm compared to a sequential version of the Aho-Corasick algorithm for DNA analysis. We may observe that the $k$-mers CoEx algorithm scales gracefully with respect to the size of data-sets and the number of threads. The maximal speedup of 17.65$\times$ is achieved for the largest data-set (that is the human DNA segment) using 48 threads. 
 
Figure \ref{fig:comparison} compares the performance of our $k$-mers CoEx algorithm with the \emph{regex-dna benchmark} \cite{REGEX_DNA}, which is implemented in C++ using the RE2 library~\cite{RE2_Library} and OpenMP. The RE2 implementation is based on splitting the pattern in smaller patterns, and matching the input string in parallel for each sub-pattern. Since the \emph{regex-dna benchmark} does not support larger data-sets, we have compared the two algorithms for the two smallest data-sets: chicken (1060MB) and turkey (193MB). Our $k$-mers CoEx algorithm outperforms the \emph{regex-dna benchmark} for both data-sets. We may observe that the $k$-mers CoEx algorithm running on one thread takes the same amount of time as the \emph{regex-dna benchmark} running on the total amount of threads. This happens because one thread has to perform at least one sequential REM for a specific sub-pattern. In this class of algorithms where the execution time is mainly dependent on the length of the input, balancing the work among the available threads should be done by splitting the input string, instead of the pattern length. One could benefit from partitioning a long pattern into smaller one, in cases when the input string is either relatively short or can not be split (Real-time Network Intrusion Detection).



\section{Related Work} 
\label{related_work}
In this section we discuss the state-of-the-art in pattern matching and DNA sequence analysis techniques for multi-core architectures.

Existing approaches use both hardware and software to accelerate the process of regular expression matching. In comparison to hardware based state machines, which are faster, less flexible and more expensive, software based acceleration techniques are flexible in terms of updating or adding new patterns \cite{SoewitoW07}.

Herath et al. presented in \cite{Herath} an implementation of the Aho-Corasick string matching algorithm using POSIX threads, which is based on the pattern partitioning approach. A replication of the Herath's study with the intention to improve the software implementation of the Aho-Corasick algorithm was conducted by Arudchutha et al. \cite{arudchutha2014string}.

Mar\c{c}ais and Kingsford \cite{MarcaisK11-jellyfish} present the Jellyfish tool, which is based on the lock-free hash table that is optimized for counting $k$-mers of length up to 31 bases. Rizk et al. \cite{RizkLC13} present a similar approach to Jellyfish \cite{MarcaisK11-jellyfish}, so called DSK, which is designed for small-memory servers. The $k$-mers are counted by traversing the hash tables. Using hash tables for the internal representation resulted to be memory inefficient \cite{DrewsLW10}. As described by Drews et al. \cite{DrewsLW10} a sequence corresponding to a human chromosome with 24-230MB of input data would require gigabytes of memory to store the $k$-mers information.

Drews et al. \cite{DrewsLW10} achieved significant speedup by partitioning the input string among the threads in such a way that each thread processes only sequences starting with a specified prefix used to divide the radix tree among the threads. They achieved up to 6.9$\times$ speedup on a shared memory system with 8 cores.

The n-step FM-index approach presented by Chac\'{o}n et al. \cite{ChaconMEH13} achieved speedups from 1.4$\times$ to 2.4$\times$ with respect to their original FM-index search algorithm. 

An approach based on the Aho-Corasick string matching algorithm designed for the Cray XMT architecture is proposed by Villa et al. \cite{VillaCM09}. They split the input among the available threads, and overlap the input by the pattern length. Their approach is applicable for multiple patterns as long as they are of the same length, otherwise, the occurrences of the shortest patterns occurring on the crossing border may be counted by both threads. 

A method for searching arbitrary regular expressions using speculation is proposed by Luchaup et al. \cite{LuchaupSEJ11}. The drawback is that if an REM performed by a thread does not converge on its sub-input, then the next thread has to start from a new state that breaks the serialization and limits the scalability.

Our $k$-mers CoEx algorithm is tailored for large-scale DNA analysis. In our approach, the DNA segment is split into several chunks, and efficient speculations of the possible initial states for each chunk are performed. Furthermore, our algorithm optimizes the REM using a converging point.


\section{Summary} \label{summary_future_work}

We have described a parallel algorithm based on Finite Automata for counting and extracting $k$-mers in a DNA segment. In a series of experiments with real world data-sets we have observed that the algorithm scales well with respect to various problem and machine sizes. We achieved the maximal speedup of 17.65$\times$ for the largest data-set (that is the human DNA segment) using 48 threads on a dual-socket shared-memory system with 24 physical cores. In comparison to the \emph{regex-dna benchmark} our algorithm was up to three times faster. 

In this paper we have studied the performance of our approach for DNA sequence analysis on a shared-memory system with two 12-core Intel Xeon processors. It may be useful to compare the performance that we achieved using all available cores of the host Intel Xeon processors with the performance achieved when all available cores of the Intel Xeon Phi coprocessor are used \cite{mp-pbio15}. Furthermore, software technologies, such as \cite{dokulil13}, enable the use of all cores of homogeneous processors of the host and all available cores of the coprocessor. 

Future work could address generalization of our approach for DNA sequence analysis for various types of accelerated systems using techniques that ensure performance portability \cite{kessler12,sandrieser12,benkner11,pllana09}. The use of modeling and simulation techniques \cite{fahringer04,pllana08,pbb08,brandic06} could help to reason about the performance on extreme-scale computing architectures \cite{abraham15}.


\end{document}